\documentclass[preprint,aps,showpacs,nofootinbib,preprintnumbers,amsmath,amssymb]{revtex4-1}
\usepackage{epsfig}
\usepackage{graphicx}
\usepackage{subfigure}
\usepackage{dcolumn}
\usepackage{bm}
\usepackage[dvips]{color}
\usepackage{slashed}

\begin{document}
\title{ Correlation of $h\rightarrow\gamma\gamma$ and $Z\gamma$ in  Type-II seesaw neutrino model}
\author{Chian-Shu Chen$^{1,2}$\footnote{chianshu@phys.sinica.edu.tw}, Chao-Qiang Geng$^{1,2}$\footnote{geng@phys.nthu.edu.tw}, Da Huang$^{1}$\footnote{dahuang@phys.nthu.edu.tw}, and Lu-Hsing Tsai$^{1}$\footnote{lhtsai@phys.nthu.edu.tw}}
\affiliation{$^{1}$Department of Physics, National Tsing Hua University, Hsinchu, Taiwan
\\$^{2}$Physics Division, National Center for Theoretical Sciences, Hsinchu, Taiwan}
\begin{abstract}
We study the charged scalar contributions to the Higgs decay channels of $h\rightarrow \gamma\gamma$ and $h\rightarrow Z\gamma$
 in the Type-II seesaw neutrino model.
In most of the allowed parameter space in the model, the new contribution to $h\rightarrow Z\gamma$
 is positively correlated with that to $h\rightarrow\gamma\gamma$.
If the current excess of the $h\rightarrow \gamma\gamma$ rate measured by the  ATLAS  Collaboration persists,
the $h\rightarrow Z\gamma$ rate should be also larger than the corresponding standard model prediction.
We demonstrate that the anti-correlation between
 $h\rightarrow\gamma\gamma$ and $h\rightarrow Z\gamma$ only exists in some special region.
\end{abstract}
\date{\today}
\maketitle
\section{Introduction}
Current experimental results at the LHC for the Higgs search are consistent with the predictions of
the Higgs boson ($h$) in the standard model (SM)~\cite{atlas:2012gk, cms:2012gu}. 
However, in the $h\rightarrow \gamma\gamma$ decay channel,
there exists some inconsistency between ATLAS and CMS Collaborations,
in which the observed rate is $1.6^{+0.3}_{-0.3}$~\cite{ATLAS_NOTE_2013_034} and 
$0.78^{+0.28}_{-0.26}$~\cite{moriond2} in comparison with the SM prediction~\cite{SM1,SM2,SM3,SM4,SM5}, respectively.
Although there is no significant discrepancy with respect to the SM in the diphoton mode at the moment,
 if the excess (deficit) seen by the ATLAS (CMS)
 is confirmed by the future measurements,
some new physics explanation is clearly needed.
 Theoretically, models
with additional charged particles in the loops are the common approaches to enhance the decay rate of
$h\rightarrow \gamma\gamma$~\cite{dipho_1,dipho_2,dipho_3,dipho_4,dipho_5,dipho_6,dipho_7,
dipho_8,dipho_9,dipho_10,dipho_11,dipho_12,dipho_13,dipho_14,dipho_15,
dipho_16,dipho_17,dipho_18,dipho_19}.
It was pointed out that a combined analysis of $h\rightarrow \gamma\gamma$ and $h\rightarrow Z\gamma$
could provide more complete electroweak charge structure of these new physics and hence, test the feasibility of these models
more precisely~\cite{Djouadi:1996yq,Djouadi:2005gi,Kniehl:1993ay,Carena:2012xa,
Chiang:2012qz,Dorsner,Picek:2012ei,Huang:2012rh,Li:2012mu,Chen,Cao}.
The  Type-II seesaw mechanism~\cite{typeII_1,typeII_2,typeII_3,typeII_4,typeII_5,typeII_6,typeII_7}  is a well-motivated way to generate small neutrino masses with additional
charged scalars beyond the SM and its related studies on $h\rightarrow\gamma\gamma$ have been devoted in
Refs.~\cite{Melfo:2011nx,Arhrib:2011vc,Akeroyd:2012ms,Wang:2012ts,Chun:2012jw}.
 The decay rate of $h \rightarrow Z\gamma$ in the Type-II seesaw model
has been recently investigated in Ref.~\cite{Dev:2013ff} and found an interesting correlation between $h\rightarrow \gamma\gamma$
and $h\rightarrow Z\gamma$ due to  the doubly charge scalar $H^{++}$. In this paper, we further consider the effects of 
the singly charge scalar $H^{+}$ on $h\rightarrow\gamma\gamma$ and $h\rightarrow Z\gamma$ in 
the Type-II seesaw model by applying the general discussion in Ref.~\cite{Chen:2013vi}.
Interestingly, we obtain a correlated relation between $h \rightarrow \gamma\gamma$ and $h \rightarrow Z\gamma$
 in most of the parameter space in the model.
 The anti-correlation between
 $h\rightarrow\gamma\gamma$ and $h\rightarrow Z\gamma$ can only exist in some
 special case.
This paper is organized as follows. In Sec.~II, we briefly introduce the Type-II seesaw model. In Sec.~III, the correlation between 
$h\rightarrow Z\gamma$ and  $h\rightarrow \gamma\gamma$ is studied.  Conclusion are given in Sec.~IV.

\section{Type-II Seesaw Model}
In the  Type-II seesaw model~\cite{typeII_1,typeII_2,typeII_3,typeII_4,typeII_5,typeII_6,typeII_7}, a scalar triplet $\Delta$ with its representation $(3,2)$
under $\mathrm{SU}(2)_L\times\mathrm{U}(1)_Y$ gauge groups
 is introduced, which can be expressed as
\begin{eqnarray}
\label{Triplet}
\Delta=
\left(\begin{array}{cc}
{1\over\sqrt{2}}\Delta^+&\Delta^{++}\\
\Delta^{0}&-{1\over\sqrt{2}}\Delta^+\\
\end{array}\right)\;,
\end{eqnarray}
leading to the Yukawa couplings
\begin{eqnarray}
Y_{ab}\overline{(L_{La})^c}(i\sigma_2)\Delta (L_{Lb})\;+\; \mathrm{h.c.}\,,
\end{eqnarray}
with the Pauli matrix $\sigma_2$ and the symmetric matrix $Y_{ab}$.
The scalar potential of the model can be in general expressed in the form
\begin{eqnarray}
V(\Phi, \Delta)&=&-m_\Phi^2(\Phi^\dagger\Phi)+\lambda(\Phi^\dagger\Phi)^2+M_\Delta^2\mathrm{Tr}(\Delta^\dagger\Delta)
+\lambda_1[\mathrm{Tr}(\Delta^\dagger\Delta)]^2+\lambda_2\mathrm{Tr}\;(\Delta^\dagger\Delta)^2\nonumber\\
&&+\lambda_3(\Phi^\dagger\Phi)\mathrm{Tr}(\Delta^\dagger\Delta)+\lambda_4\Phi^\dagger[\Delta^\dagger\,,\,\Delta]\Phi+
\left({\mu\over\sqrt{2}} \Phi^T i\sigma_2 \Delta^\dagger \Phi +\mathrm{h.c.} \right)\;,
\label{potential}
\end{eqnarray}
where $\Phi$ is the SM Higgs doublet with the vacuum expectation value (VEV) $\langle\Phi\rangle=(0,\, v/\sqrt{2})^T$,
and all parameters in the potential are taken to be real without loss of generality.
Note that the potential in Eq.~(\ref{potential}) becomes to the one in Ref.~\cite{Dev:2013ff}
 after the use of  the transformations: $\lambda\rightarrow \lambda/2$, $\lambda_1\rightarrow (\lambda_1+\lambda_2)/2$,
 $\lambda_2\rightarrow-\lambda_2/2 $, $\lambda_3\rightarrow \lambda_4$, $\lambda_4\rightarrow \lambda_5$, and $\mu\rightarrow \Lambda_6$.
The neutral component  $\Delta^0$ of the triplet scalar in Eq.~(\ref{Triplet})
acquires its VEV $v_\Delta=\sqrt{2}\langle \Delta^0\rangle$ through the relation
\begin{eqnarray}
v_\Delta[2M_\Delta^2 +(\lambda_3-\lambda_4){v^2}+2(\lambda_1+\lambda_2)v_\Delta^2]-\mu v^2=0\;,
\end{eqnarray}
where $M_{\Delta}$ represents the mass scale of the triplet scalar. For the case of $M_{\Delta} \gg v$,
such as the grand unification scale,
the triplet VEV is naturally suppressed as $v_{\Delta} \approx \mu v^{2}/(2M^2_{\Delta})$. In this scenario, the extra scalars will have
no significant effects on the collider phenomena. In this paper, we concentrate on the mass scale  where the triplet $\Delta$ is testable within the LHC
search. In this case, we expect $M_{\Delta} \approx v$ so that $v_{\Delta} \sim \mu$.
On the other hand,
$v_\Delta$ is constrained to have an
upper bound $v_\Delta\lesssim {\cal O}(1)~\mathrm{GeV}$ by the parameter
$\rho\equiv m_W^2/ (m_Z^2\cos^2\theta_W)= 1.004^{+0.0003}_{-0.0004}$~\cite{pdg}.
 As a result,
$v_\Delta$ comes from the nonzero coefficient $\mu$ of the last term in Eq.~(\ref{potential}),
corresponding to the breaking of lepton number symmetry. It will generate the Majorana
neutrino mass at the tree level
\begin{eqnarray}
(M_\nu)_{ab}=\sqrt{2}v_\Delta Y_{ab}\,,
\end{eqnarray}
where $a,b=e,\,\mu$ and $\tau$.
To understand the small neutrino masses, the upper bound $v_\Delta\approx 1~\mathrm{GeV}$ corresponds to
a suppressed Yukawa coupling of $Y\lesssim 10^{-9}$, whereas
 the lower bound $v_{\Delta} \approx 10^{-9}$~GeV is set if
 $Y={\cal O}(1)$.
Back to the scalar sector, the mass spectra of the scalars can be solved from Eq.~(\ref{potential}),  given by
\begin{eqnarray}
m_h^2&=&{1\over2}\left(M_{11}^2+M_{22}^2-\sqrt{(M_{11}^2-M_{22}^2)^2+4M_{12}^4}\right)\;,
\\
m_{H^0}^2&=&{1\over2}\left(M_{11}^2+M_{22}^2+\sqrt{(M_{11}^2-M_{22}^2)^2+4M_{12}^4}\right)\;,
\\
m_{A^0}^2&=&\left[M_\Delta^2+{1\over2}(\lambda_3-\lambda_4)v^2+(\lambda_1+\lambda_2) v_\Delta^2\right]\left(1+{4v_\Delta^2\over v^2}\right)\,,
\\
m_{H^{\pm}}^2&=&\left[M_\Delta^2+{1\over2}\lambda_3v^2+(\lambda_1+\lambda_2)v_\Delta^2\right]\left(1+{2v_\Delta^2\over v^2}\right)\,,
\\
m_{H^{\pm\pm}}^2&=&M_\Delta^2+{1\over2}\left(\lambda_3+\lambda_4\right)v^2+\lambda_1v_\Delta^2\;,
\end{eqnarray}
where $h$ is the SM-like Higgs,  $H^0$ and $A^0$ are the CP even and odd neutral components,
and $H^+$ and $H^{++}$ are the singly and doubly charge mass eigenstates, respectively, while the neutral scalar mass matrix elements are
\begin{eqnarray}
M_{11}^2&=&2\lambda v^2\;,\;M_{22}^2=M_\Delta^2+{1\over2}(\lambda_3-\lambda_4)v^2+3(\lambda_1+\lambda_2) v_\Delta^2\;,
\nonumber\\
M_{12}^2&=&-{2v_\Delta\over v}\left[M_\Delta^2+(\lambda_1+\lambda_2)v_\Delta^2\right]\;.
\end{eqnarray}
The mixing angles of the singly charged  and neutral scalars are approximately proportional to $v_\Delta/v$,
so the charged mass eigenstates $H^{+}$ and $H^{++}$ nearly coincide with  the weak eigenstates $\Delta^+$ ($I_3=0, Q=1$) and $\Delta^{++}$ ($I_3=1, Q=2$), respectively.
For this reason,  we will ignore the contributions from $v_\Delta$ from now on.
It is also worth noticing that the trilinear couplings for the charged scalars with the SM-like Higgs $h$ are given by
\begin{eqnarray}
\mu_{hH^+H^-}&=&\lambda_3v={2\over v}\left(m_{H^{+}}^2-M_\Delta^2\right)\,,
\label{tri1}\\
\mu_{hH^{++}H^{--}}&=&(\lambda_3+\lambda_4)v={2\over v}\left(m_{H^{++}}^2-M_\Delta^2\right)\,.
\label{tri2}
\end{eqnarray}
 From the above relations,
the deviations of the charged scalars with the triplet bare mass $M_\Delta$ clearly affect
both signs and magnitudes of the corresponding trilinear couplings. In general, the mass splitting or the gauge quantum number
of a scalar multiplet beyond the SM is also constrained by the oblique parameters. In the Type-II seesaw model, one can set the upper
bound on the mass splitting of the triplet to be $|m_{H^{++}}-m_{H^+}|\lesssim40~\mathrm{GeV}$, which is insensitive to the
triplet scale $M_\Delta$~\cite{Chun:2012jw}. The constraints for the parameters in  the scalar potential can obtained from
the  stable conditions, given by
\begin{eqnarray}
\label{condition}
\lambda\geq 0\;,\;\lambda_1+\lambda_2\geq0\;&,&\;2\lambda_1+\lambda_2\geq0\;,\nonumber\\
\lambda_3\pm\lambda_4+2\sqrt{\lambda(\lambda_1+\lambda_2)}\geq 0\;&,&\;\lambda_3\pm\lambda_4
+2\sqrt{\lambda(\lambda_1+\lambda_2/2)}\geq 0\;.
\end{eqnarray}
To ensure the perturbativity and the conditions in Eq.~(\ref{condition})  from the electroweak to
higher energy scale (e.g. Planck scale),
a positive value of $\lambda_3$ is preferred if $\lambda_{1,2}$ are taken to be small~\cite{Chun:2012jw,Dev:2013ff}.
However, a negative value of $\lambda_3$ is still possible as long as the square roots in Eq.~(\ref{condition}) are
large enough~\cite{Arhrib:2011vc,Chun:2012jw,Arhrib:2011uy}. In what follows we consider the implications of $h\rightarrow \gamma\gamma$
and $h\rightarrow Z\gamma$ in these two parameter regions.

\section{Correlation between $h\rightarrow\gamma\gamma$ and $h\rightarrow Z\gamma$}
 The general formulae for scalar ($s$), $t$-quark, and $W$-boson contributions to the decay rates of 
$h\rightarrow \gamma\gamma$ and $h\rightarrow Z\gamma$ can be derived by the Feynman rules listed in Ref.~\cite{Denner:1991kt}, and the results are given by~\cite{Chen:2013vi}
\begin{eqnarray}
\Gamma(h\rightarrow \gamma\gamma)&=&{G_F\alpha^2 m_h^3\over 128\sqrt{2}\pi^3}\left|3Q_t^2A^{\gamma\gamma}_{1/2}(\tau_t)+A^{\gamma\gamma}_1(\tau_W)
+Q_s^2{v\mu_{hss^*}\over 2m_{s}^2}A^{\gamma\gamma}_0(\tau_{s})\right|^2\,,
\label{generalformulae1}\\
\Gamma(h\to Z\gamma)&=& \frac{\alpha^2}{512\pi^3}m_h^3\left(1-\frac{m_Z^2}{m_h^2}\right)^3\left|{\cal A}_\mathrm{SM}^{Z\gamma}
- \frac{\mu_{hss^*}}{m_s^2s_Wc_W}(2Q_sQ_s^Z)
A_0^{Z\gamma}(\tau_s,\lambda_s)\right|^2\,,\label{generalformulae2}
\end{eqnarray}
with
\begin{equation}
\quad\quad\quad {\cal A}_\mathrm{SM}^{Z\gamma}= \frac{2}{v}\left[\cot\theta_W A_1^{Z\gamma}(\tau_W,\lambda_W)
+ \frac{(6Q_t)(I^3_t-2Q_t s_W^2)}{s_W c_W}A_{1/2}^{Z\gamma}(\tau_t,\lambda_t)\right]\,,
\end{equation}
where $s_W\equiv \sin\theta_W$, $c_W\equiv \cos\theta_W$, 
$\mu_{hss^*}$ is the trilinear coupling derived from the scalar potential,
$\tau_i=4m_i^2/m_h^2$, $\lambda_i=4m_i^2/m_Z^2$ ($i=W,t,s$).
 $Q_{t,s}$ are the electric charges of $t$-quark and the scalar, 
 and
$Q_{s}^Z=I_{s}^3-Q_{s}\sin^2\theta_W$
with $I_{t,s}^3$ being the third isospin components of $t$-quark and the scalar, respectively.
The loop functions $A^{\gamma\gamma}_{(0,\,1/2,\,1)}$ and $A^{Z\gamma}_{(0,\,1/2,\,1)}$ in Eqs.~(\ref{generalformulae1})
and (\ref{generalformulae2})
are defined as
\begin{eqnarray}
A_0^{\gamma\gamma} (x) &=& -x^2[x^{-1}-f(x^{-1})]\,,
\nonumber\\
A_{1/2}^{\gamma\gamma} (x) &=& 2x^2[x^{-1}+(x^{-1}-1)f(x^{-1})]\,,
\nonumber\\
A_1^{\gamma\gamma}(x) &=& -x^2[2x^{-2}+3x^{-1}+3(2x^{-1}-1)f(x^{-1})]\,,
\nonumber\\
A_0^{Z\gamma}(x,y) &=& I_1(x,y)\,, 
\nonumber\\
A_{1/2}^{Z\gamma} (x,y) &=& I_1(x,y)-I_2(x,y)\,,
\nonumber\\
A_1^{Z\gamma}(x,y) &=& 4(3-\tan^2\theta_W)I_2(x,y)+[(1+2x^{-1})\tan^2 \theta_W-(5+2x^{-1})]I_1(x,y)\,,
\end{eqnarray}
where
\begin{eqnarray}
I_1(x,y) &=& \frac{x y}{2(x-y)}+\frac{x^2 y^2}{2(x-y)^2}[f(x^{-1})-f(y^{-1})]+\frac{x^2 y}{(x-y)^2}[g(x^{-1})-g(y^{-1})]\,,
\nonumber\\
I_2(x,y) &=& -\frac{x y}{2(x-y)}[f(x^{-1})-f(y^{-1})]\;,
\end{eqnarray}
with the functions $f(x)$ and $g(x)$ in the range $x<1$, given by
\begin{eqnarray}
f(x) &=&(\sin^{-1}\sqrt{x})^2\,,
\nonumber\\
g(x)&=&\sqrt{x^{-1}-1}(\sin^{-1}\sqrt{x})\,.
\end{eqnarray}
 In the SM, the $W$-boson contributions to $h\rightarrow \gamma\gamma$ and $h\rightarrow Z\gamma$ dominate over those from $t$-quark,  
while
the signs of the corresponding amplitudes $A^{\gamma\gamma}_1$ and $A^{Z\gamma}_1$ are opposite.
The new contributions to $h\rightarrow\gamma\gamma$ or $h\rightarrow Z\gamma$ beyond the SM are usually characterized by the
expressions
\begin{eqnarray}
R_{\gamma\gamma(Z\gamma)}={\sigma(pp\rightarrow h)\mathrm{Br}(h\rightarrow\gamma\gamma(Z\gamma))
\over \sigma_\mathrm{SM}(pp\rightarrow h)\mathrm{Br}_\mathrm{SM}(h\rightarrow\gamma\gamma(Z\gamma))}\;.
\end{eqnarray}
In our case, the SM-like Higgs production rates are almost the same as those for the SM since the mixing with the triplet is very small.
For  $h\rightarrow \gamma\gamma$,  the type of the interference between
 the new charged scalar and the SM contributions only depends on
the sign of $\mu_{hss^*}$ since $Q_s^2$ is always positive. In our discussion, the trilinear couplings of $H^{+}$ and $H^{++}$ to
$h$ are given in Eqs.~(\ref{tri1}) and (\ref{tri2}). If $\mu_{hss^*}$ is negative (positive),
then the interference with the SM one is constructive
(destructive).
The situation in $h\rightarrow Z\gamma$ is more
complicated~\cite{Chen:2013vi}. To determine whether the new charged scalar contribution to $h\rightarrow Z\gamma$ is
constructive or destructive, we need to know  the sign of not only $\mu_{hss^*}$, but also the charge combination
$Q_sQ_s^Z=(I_3+Y/2)(I_3\cos^2\theta_W-Y\sin^2\theta_W/2)$.
It is obvious that a larger value of $I_3$ yields a positive value of $Q_sQ_s^Z$,
 whereas $Q_sQ_s^Z$ becomes negative for a larger $Y$. 
 Finally, by comparing with the SM amplitudes, it is easy to see from Eqs.~(\ref{generalformulae1}) and
 (\ref{generalformulae2}) that a positive (negative) value of $Q_sQ_s^Z$ will lead to the correlated (anti-correlated) behavior 
 between $h\rightarrow\gamma\gamma$ and $h\rightarrow Z\gamma$.\footnote{We note that our result in Eq.~(\ref{generalformulae2})
 is different from Eq.~(5.11) in Ref.~\cite{Dev:2013ff}.
 The reason for the difference  lies in that the scalar contribution to $h\to Z\gamma$ in Eq.~(5.11) used by the authors of
  Ref.~\cite{Dev:2013ff} has an extra factor $-1/(2\sin^2\theta_W)$~\cite{Referee}.}

\begin{figure}
  \centering
  \includegraphics[width=8cm]{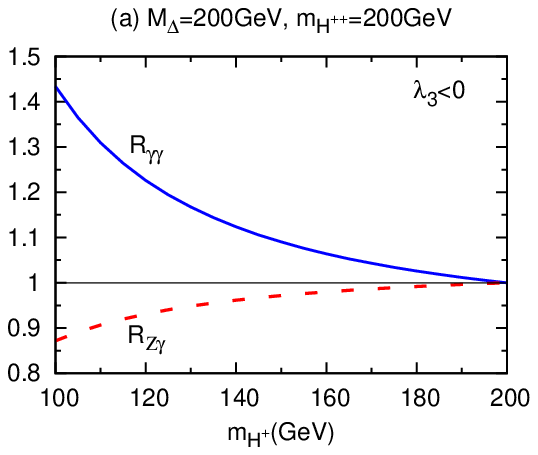}
  \includegraphics[width=8cm]{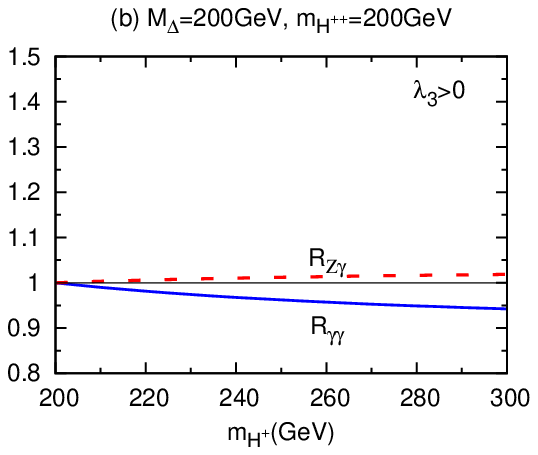}
  \caption{$R_{\gamma\gamma}$ (solid) and $R_{Z\gamma}$ (dashed) versus $m_{H^+}$ with
  $m_{H^{++}}=M_\Delta=200$ GeV
  for (a) $m_{H^{+}}<M_{\Delta}\,(\lambda_3<0)$ and
  (b) $m_{H^{+}}>M_{\Delta}\,(\lambda_3>0)$.}
  \label{mHp}
\end{figure}

In the  Type-II seesaw model, new contributions to $h\rightarrow \gamma\gamma$ and $h\rightarrow Z\gamma$
arise only from the loops involving with the charged scalars of $H^{\pm}$ and $H^{\pm\pm}$.
Since the mixing between the doublet and triplet scalars is ignored,
$Q_sQ_s^Z$ are negative and positive for $H^{+}$ and $H^{++}$
as they approximately correspond to $I_3=0$ and $ 1$, respectively.
Clearly, for having ${H^+}$ alone, the rates of
$h\rightarrow \gamma\gamma$ and
$h\rightarrow Z\gamma$  are anti-correlated.
We may set $m_{H^{++}} = M_\Delta$ to eliminate the contributions of $H^{++}$
and plot with $M_\Delta=200$~GeV as presented in Fig.~\ref{mHp}.
The anti-correlated region with $m_{H^+}<M_\Delta$,
corresponding to $\lambda_3<0$, is shown in Fig.~\ref{mHp}a.
We note that this parameter space allowed by the
constraints from the vacuum stability and oblique parameters is small.
On the other hand, for $m_{H^+}>M_\Delta$ with $\lambda_3>0$,
the results are depicted  in Fig.~\ref{mHp}b. In this case,
the $H^{+}$ domination is not preferred as  the $h\rightarrow \gamma\gamma$ rate gets reduced, which
conflicts with the current data at the LHC.

\begin{figure}
  \centering
  \includegraphics[width=8cm]{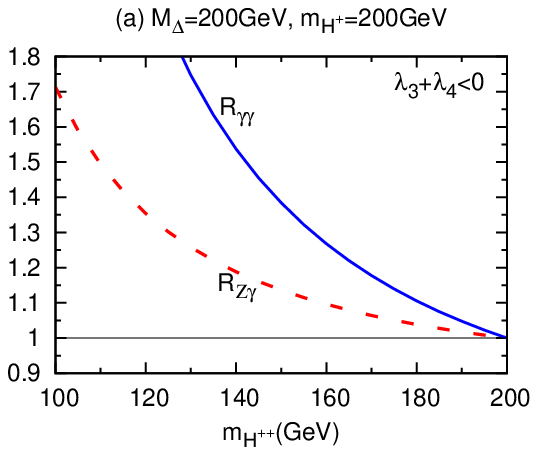}
  \includegraphics[width=8cm]{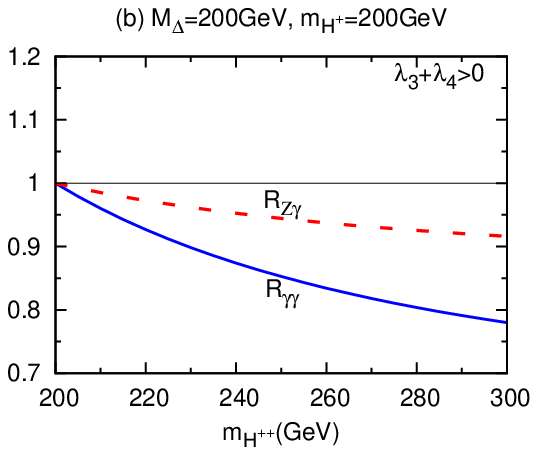}
  \caption{$R_{\gamma\gamma}$ (solid) and $R_{Z\gamma}$ (dashed) versus $m_{H^{++}}$ with
  $m_{H^{+}}=M_\Delta=200$ GeV
  for (a) $m_{H^{++}}<M_{\Delta}\,(\lambda_3+\lambda_4<0)$ and
  (b) $m_{H^{++}}>M_{\Delta}\,(\lambda_3+\lambda_4>0)$.}
  \label{mHpp}
\end{figure}

\begin{figure}
  \centering
  \includegraphics[width=8cm]{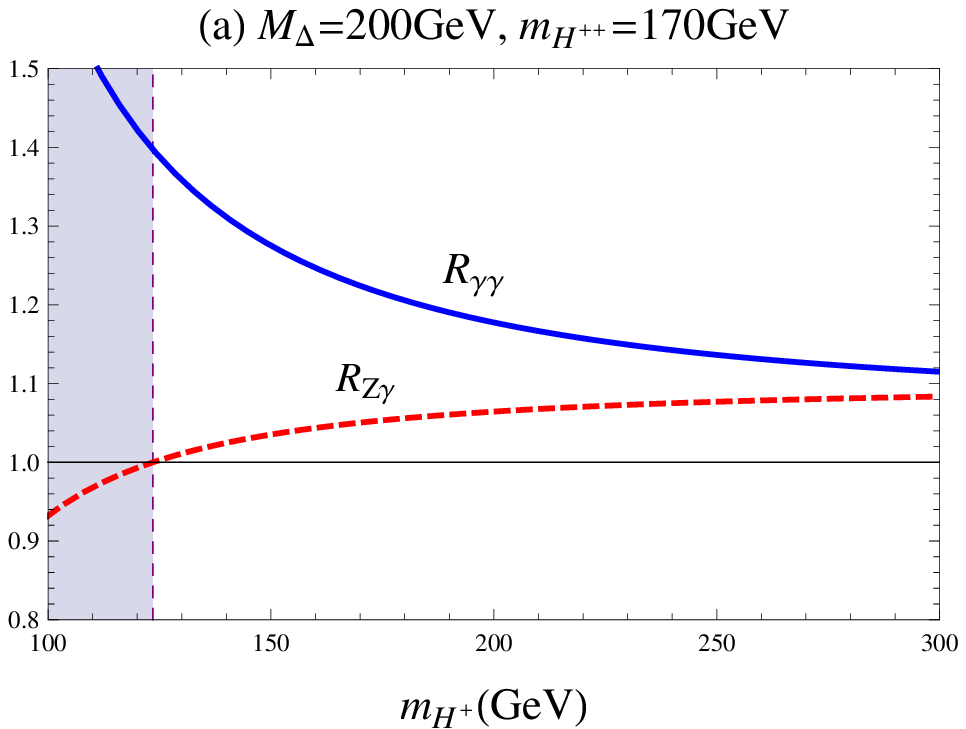}
  \includegraphics[width=8cm]{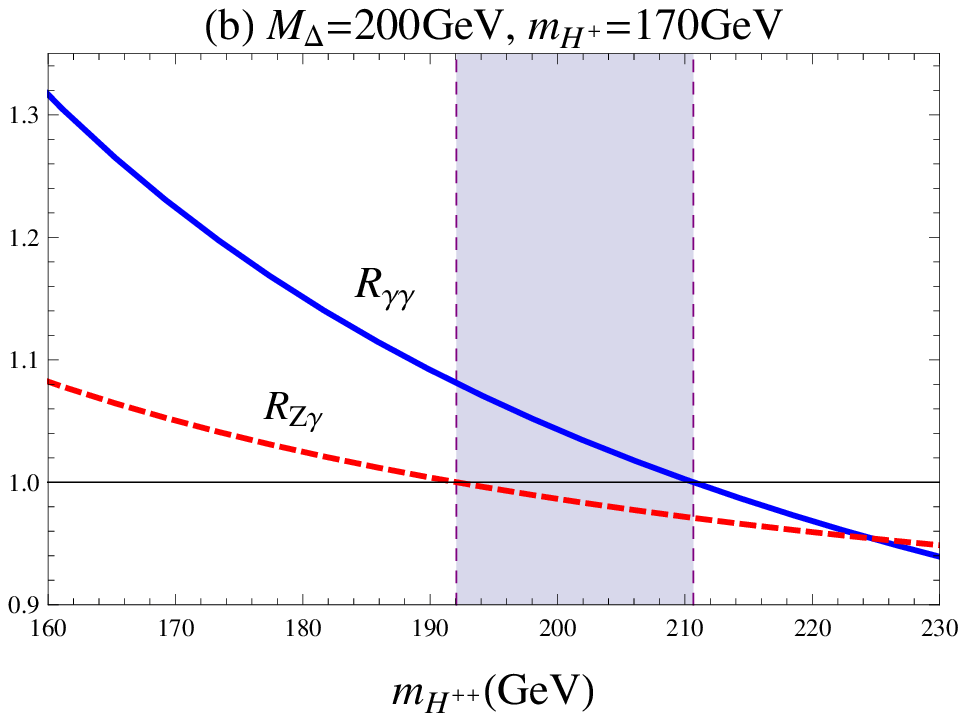}
  \caption{$R_{\gamma\gamma}$ (solid) and $R_{Z\gamma}$ (dashed) versus (a) $m_{H^+}$
  with $m_{H^{++}}=170$ GeV and (b) $m_{H^{++}}$ with  $m_{H^{+}}=170$ GeV,
  where  $M_\Delta=200$ GeV and
 the  shaded areas represent the  anti-correlated regions.}
 \label{CombinedResults}
\end{figure}

In Fig.~\ref{mHpp},
we give the related decay rates for the new contributions only from $H^{++}$, which is
equivalent to set $m_{H^{+}} = M_\Delta$. In this case, the rates of
$h\rightarrow\gamma\gamma$ and $h\rightarrow Z\gamma$ are correlated with each other as shown in Fig.~\ref{mHpp}.
Similarly, the region with
$\lambda_3+\lambda_4>0$ is not preferred by the LHC results.
It is important to note that for $\lambda_3+\lambda_4<0$,
the constraint on the mass difference between $m_{H^+}$ and $m_{H^{++}}$ from the oblique parameters
also  limits the value of $m_{H^{++}}$,
so that the $h\rightarrow \gamma\gamma$ rate can not be arbitrarily large.
Finally, in Fig.~\ref{CombinedResults}
 we illustrate the general case with both $H^+$ and $H^{++}$ contributions being taken
into account, where we have fixed the masses of $H^{++}$ and $H^{+}$ to be
$170$ GeV in Figs.~\ref{CombinedResults}a and \ref{CombinedResults}b, respectively.
It turns out that in most of the allowed parameter space,
the $H^{++}$ contributions are dominant, resulting in the positive correlation
between the $h\rightarrow\gamma\gamma$ and $h\rightarrow Z\gamma$ rates,
  since both $Q_s^2$ and $Q_sQ_s^Z$ are larger than
those of $H^{+}$. 
 It is also consistent with the results shown in Ref.~\cite{Chen:2013vi}.
 However, the anti-correlation can still exist if the $H^{+}$ contributions dominate over those from
 $H^{++}$. For example, one can enhance the $H^{+}$ contributions
 by reducing  $m_{H^{+}}$ and  increasing $\mu_{hH^{+}H^{+}}$ simultaneously, as
 plotted in Fig.~\ref{CombinedResults}a with $m_{H^+}\lesssim125$ GeV.
Another way is to suppress the $H^{++}$ contributions by setting $M_{\Delta} \approx m_{H^{++}}$
as in the region $190~\mathrm{GeV}\lesssim m_{H^{++}}\lesssim210~\mathrm{GeV}$ in Fig.~\ref{CombinedResults}b.

\section{Conclusions}
 We have studied the details of $h\rightarrow \gamma\gamma$ and $h\rightarrow Z\gamma$ rates in the Type-II seesaw model.
In particular, we have shown that the contributions to $h\rightarrow \gamma\gamma$ and $h\rightarrow Z\gamma$
from $H^{++}$ ($H^{+}$) by itself
 are (anti-)correlated.
On the other hand, for the general case with the existences of
both $H^{+}$ and $H^{++}$,
we have found that the deviation of the $h\rightarrow Z\gamma$ rate from the SM prediction
has the same sign as the $h\rightarrow \gamma\gamma$ counterpart in most of the parameter space,
whereas in some small regions with $\lambda_3<0$ and $m_{H^{++}}\simeq M_\Delta$,
the anti-correlation between $h\rightarrow \gamma\gamma$ and $h\rightarrow Z\gamma$ appears, which could  be tested in the
future experiments at the LHC.

\begin{acknowledgments}
The work was supported in part by National Center for Theoretical Science, National Science
Council (NSC-98-2112-M-007-008-MY3 and NSC-101-2112-M-007-006-MY3) and National Tsing-Hua
University (102N1087E1 and 102N2725E1), Taiwan, R.O.C.
\end{acknowledgments}

\end{document}